\documentclass[epjCONF,columns]{svjour} 
\usepackage{graphicx}
\usepackage[varg]{txfonts} 
\usepackage[latin1]{inputenc}
\session-title{2011 Hadron Collider Physics symposium (HCP-2011)}
\begin{document}
\title{Measurement of the ratio of branching fractions
			$\frac{\mathcal{BR}(B^0\to K^*\gamma)}{\mathcal{BR}(B_s^0\to\phi\gamma)}$}
\author{Savrina Daria\inst{1,2} on behalf of the LHCb collaboration\fnmsep\thanks{\email{Daria.Savrina@cern.ch}}}
\institute{Institute of Theoretical and Experimental Physics (ITEP, Moscow) \and
Lomonosov Moscow State University (SINP MSU)}
\abstract{
The interest to the rare radiative decays of the B-mesons at LHCb is mostly aroused due to the measurement of the photon polarization in the $B_s^0\to\phi\gamma$ decay, which may provide a sensitive probe for the Standard Model. The LHCb experiment has started to take data at the energy of $\sqrt{s} = 7~TeV$ in 2010 and the current paper presents the result of the studies of the two rare radiative decays $B^0\to K^*\gamma$ and $B_s^0\to\phi\gamma$ with $340 pb^{-1}$ of data taken in the first half of 2011. With this data we have a preliminary measurement of $\frac{\mathcal{BR}(B^0\to K^*\gamma)}{\mathcal{BR}(B_s^0\to\phi\gamma)} = 1.52\pm0.14(stat.)\pm0.10(syst.)\pm0.12(f_s/f_d)$ and assuming the measured value of $\mathcal{B}(B^0\to K^*\gamma) = 	( 4.33 \pm 0.15 )\times10^{-5}$ we infer $\mathcal{B}(B_s^0\to \phi\gamma) = (2.8\pm0.5)\times10^{-5}$ \cite{bib:conf}.
} 
\maketitle
\section{Introduction}
\label{intro}
Rare radiative decays of the B-mesons provide a good test for the Standard Model (SM). Being forbidden at tree level, such processes may only occur due to loop diagrams involving FCNC and thus become very sensitive to the impact of non-standard particles. The accuracy of the theoretical predictions for such decays makes them attractive from the experimental point of view.

The theoretically predicted branching fractions have the same value of $(4.3\pm1.4)\times10^{-5}$ for both decays \cite{bib:theo}. The experimental observations were done by the Belle (both $B^0\to K^*\gamma$ and $B_s^0\to\phi\gamma$ decays), BaBar and CLEO ($B^0\to K^*\gamma$ only) collaborations, which found a good agreement between the measured branching fractions of these decays and the SM predictions (see Table \ref{tab:brexp}).
\begin{table}[h]
\caption{Experimental measurements for $B^0\to K^*\gamma$ and $B_s^0\to\phi\gamma$ branching fractions \cite{bib:exp}.}
\label{tab:brexp}       
\begin{tabular}{ccc}
\hline\noalign{\smallskip}
\ & $B^0\to K^*\gamma$ & $B_s^0\to\phi\gamma$  \\
\noalign{\smallskip}\hline\noalign{\smallskip}
BaBar ($10^{-5}$) & $4.47\pm0.10\pm0.16$ &  \\
Belle ($10^{-5}$) & $4.01\pm0.21\pm0.17$ & $5.7^{+1.8+1.2}_{-1.5-1.1}$ \\
CLEO ($10^{-5}$) & $4.55^{+1.8+1.2}_{-1.5-1.1}\pm0.34$ &  \\
\noalign{\smallskip}\hline
\end{tabular}
\end{table}

The effects of the New Physics in these decays may nevertheless be discovered through the other observables, like for example $~CP$ and isospin asymmetries. The measurement of the photon polarization in the $B_s^0\to\phi\gamma$ decay, which is expected to be sensitive to the Left-Right Symmetric Model or unconstrained MSSM, is marked as one of the key measurements of the LHCb experiment. In the SM, the photons emitted in the $b\to s\gamma$ transitions should be predominantly left-handed and the emission of the right-handed photons is suppressed with a factor of $m_s/m_b$ \cite{bib:pol}. Any excess of the "wrongly"-polarized photons is considered as the evidence of the New Physics.

Here we present the result of the most precise measurement of the branching fractions ratio of the two rare radiative decays, which is also the first step towards many other exciting measurements.

\section{The LHCb detector}
\label{sec:detector}
LHCb detector is a single arm forward spectrometer covering 300 mrad solid angle, which in terms of rapidity corresponds to the $1.9 < \eta < 4.9$ region \cite{bib:det}. The acceptance for the B-mesons is estimated to be $75\ \mu barn$. The detector is constructed of the following parts:
\begin{itemize}
	\item Vertex Locator (VELO), which is located around the beam collision point. It allows to measure the positions of primary and secondary vertices with the spatial resolution of $13~\mu m$ (x-, y- directions) and $70~\mu m$ (z-direction) and propertime resolution of $50~fs$
	\item two RICH detectors, providing the pion/kaon separation with PID efficiency of more than $90\%$ and less than 10\% misidentification rate in the $2-100~GeV/c$ momentum region
	\item tracking system
	\item calorimeter system, consisting of four subsystems (scintillator pad detector SPD, preshower detector PS, electromagnetic calorimeter ECAL and hadron calorimeter HCAL), which perform the photon/electron/$\pi^0$ separation and measure the energies and positions of these particles
	\item muon system, providing ~$97\%$ effciency for the muon identification and misidentification rate below 1\% \cite{bib:stat}.
\end{itemize}
LHCb has started to take data at the center of mass energy of $\sqrt{s} = 7~TeV$ since 2010. The preliminary results shown at this conference are obtained with $340~pb^{-1}$ of data taken in the first half of 2011.

\section{Electromagnetic calorimeter calibration}
\label{sec:calib}
A good calibration of the electromagnetic calorimeter is extremely important for the 
studies of the radiative decays. Monte-Carlo simulation shows that the $3\%$ miscalibration of the 
ECAL leads to almost $20\%$ increase in the B-meson mass resolution.

The LHCb ECAL consists of 6016 modules \cite{bib:calo}, built under shashlik sampling technology, consisting of the alternating scintillating tiles ($4~mm$ thick each) and lead plates ($2~mm$ thick) and having the designed energy resolution of $\frac{\sigma_E}{E} = \frac{10\%}{\sqrt{E}}\oplus1\%$. The energy resolution of each module has been determined at the test beam.

The light from each module is collected by the wavelength shifting fibers (WLS) and transmitted to the photoelectron multipliers (PMT). The PMT gains are determined with the help of LED system installed in the calorimeter.

For the further improvement of the performance a combination of several "in situ" methods has been used. First, the ECAL is precalibrated with the help of the Energy Flow method, which is based on the idea that the distribution of the transverse energy over the surface of the calorimeter should be a smooth function of coordinates and allows to achieve $4\%$ level in cell intercalibration.

Then, for the final calibration the "Mass distribution fit" method is used. It is based on the measurement of a well known value, namely the mass of a resolved $\pi^0$-meson in its decay into two photons, relying on the calorimeter information only and aims to achieve less than $2\%$ fine calibration accuracy.

The whole calibration chain helps to improve the resolution for many particles in their decays with photons in the final states.

\section{Event selection}
\label{sec:sel}
In order to reduce the systematic errors on the measured ratio of branching fractions, 
the selections for two decays were made as close to each other as possible, thus making 
use of the common topology, kinematics and photon efficiency of the decays.

\subsection{Trigger selection}
\label{ssec:trig}
LHCb trigger system is aimed to reduce the incoming amount of data from detector from $20~MHz$ level to $3 kHz$, which is written to the tape. It includes two levels: a purely hardware level-zero trigger (L0) and high level trigger (HLT) represented by a number of C++ algorithms. The L0 trigger selects the events containing high transverse momentum photon, electron, hadron or muon candidate and passes them to the HLT. The latter one is divided into two stages: HLT1 and HLT2. Having an additional information from VELO and tracking system the HLT1 confirms or rejects the L0 decision, reducing the rate and the HLT2, having access to the information from most of the detectors, runs exclusive or inclusive selections on the almost totally reconstructed events.

Radiative decays are triggered by the L0 trigger as containing a high transverse momentum photon candidate and at the HLT1 step an additional information from the tracking system is used to select events with a high momentum kaon or pion candidate and HLT2 contains two dedicated exclusive selection lines for the $B^0\to K^*\gamma$ and $B_s^0\to\phi\gamma$ decays.

\subsection{Offline selection}
\label{sec:offl}
A pair of charged tracks of a good quality originating from the same vertex are used to construct a vector meson ($K^*$ or $\phi$). Each of the tracks is identified to be a kaon or a pion with the help of logarithmic likelihood technique, based on the information from different detector subsystems. Both tracks are required not to point to the primary vertex (track impact parameter $\chi^2 > 25$) and have a transverse momentum in excess of $500~MeV/c$. An unconstrained vertex fit is applied to the common vertex of the two tracks and the quality of this fit is also used as the selection criterion (vertex $\chi^2 < 9$). Mass of the $\phi$ candidate should not be too far  from the known PDG value ($\Delta m_{\phi} < 10~MeV/c^2$), while the mass window of the $K^*$ meson is taken equal to its natural width ($\Delta m_{K^*} < 55~MeV/c^2$).

Photon is reconstructed as an energy cluster in the electromagnetic calorimeter, having no matching track. Additional information from the SPD and Prs detectors is used for the photon identification. The transverse energy of the photons is required to exceed $2600~MeV$ and the contribution from the merged neutral pions is rejected with the help of the special photon/pion separation tool based on the shape of the electromagnetic shower in the ECAL.

Finally, the photon candidate is combined with the selected $K^*$ or $\phi$ meson to form a $B$-meson candidate. The $B$-candidate is required to have high transverse momentum ($P_T(B) > 3~GeV/c$) and point to the primary vertex with the impact parameter $\chi^2 < 9$.

\begin{figure}[t]
\centering
\includegraphics[width=80mm]{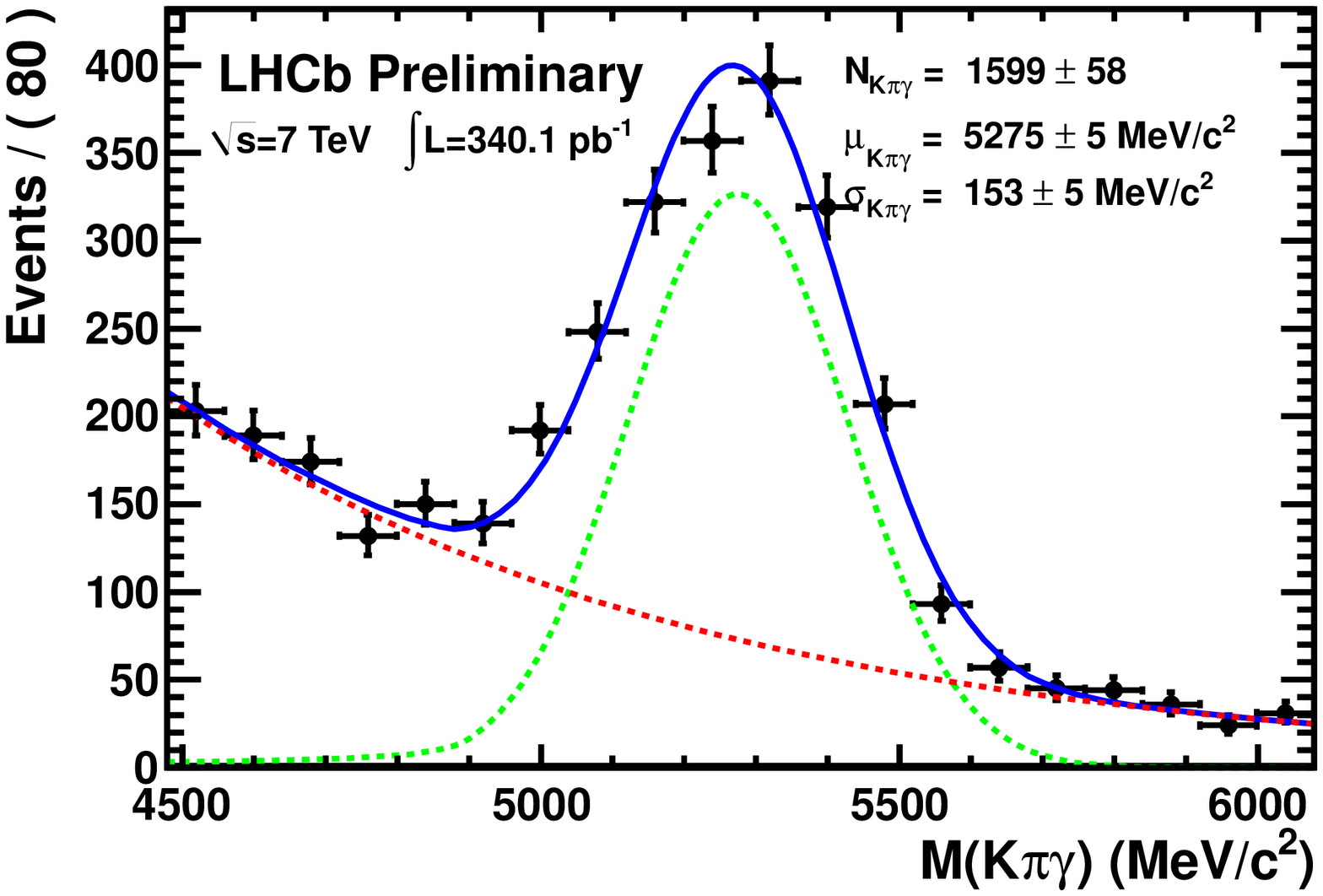}
\put(-180,110){a)}\\
\includegraphics[width=80mm]{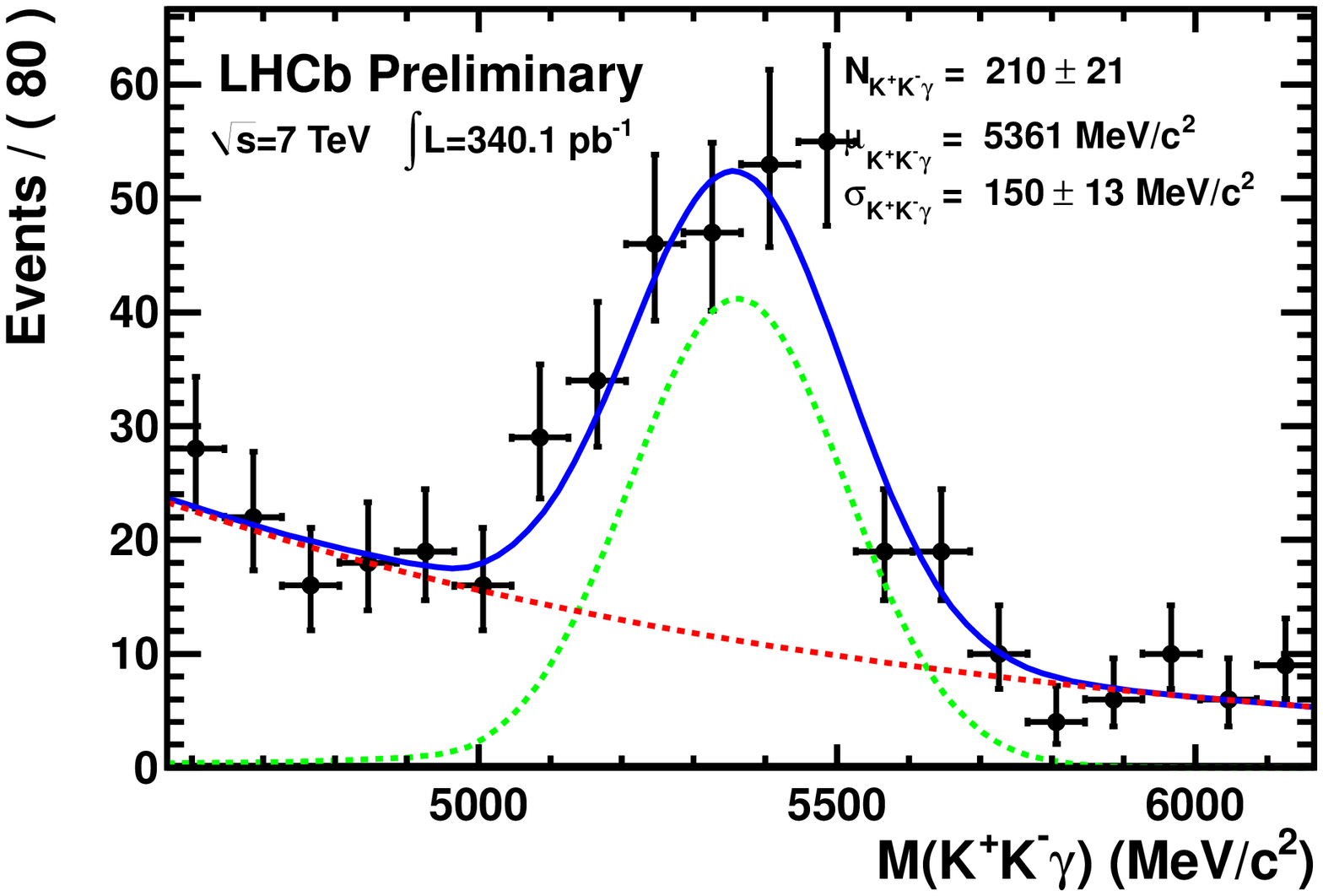}
\put(-180,110){b)}
\caption{Invariant mass distributions for a) $K^{\pm}\pi^{\mp}\gamma$ and b) $K^{+}K^{-}\gamma$ combinations with the $B^0$ and $B_s^0$ peaks visible.}
\label{fig:peaks}       
\end{figure}

\section{Measurement of the branching fraction ratio}
\label{sec:ratio}

As the result of the selection described in Sec.\ref{sec:sel} the two rare radiative decays: $B^0\to K^*\gamma$ and $B_s^0\to\phi\gamma$ were observed. The invariant mass distributions are presented in Fig.\ref{fig:peaks}, the distributions are described with a sum of Crystal Ball (signal) and exponential (background) functions. The fit is performed simultaneously for the both peaks with the mass difference fixed at the known PDG value. The signal to background ratio in $\pm3\sigma$ region is estimated to be at the level of 1.7 and the number of the $B_s^0\to\phi\gamma$ events in the peak is currently the world's largest sample of $B_s^0$ decays in this channel.

The branching fractions ratio then was defined as the following:
$$\frac{\mathcal{B}(B^0\to K^{*0}\gamma)}{\mathcal{B}(B_s^0\to \phi\gamma)} = 
\frac{\mathcal{Y}_{B^0\to K^{*0}\gamma}}{\mathcal{Y}_{B_s^0\to \phi\gamma}}\times
\frac{\mathcal{B}(\phi\to K^{+}K^{-})}{\mathcal{B}(K^{*0}\to K^{\pm}\pi^{\pm})}\times
\frac{f_s}{f_d}\times
\frac{\varepsilon_{B_s^0\to \phi\gamma}}{\varepsilon_{B^0\to K^{*0}\gamma}},$$
where $\mathcal{Y}_{B^0\to K^{*0}\gamma}$ and $\mathcal{Y}_{B_s^0\to \phi\gamma}$ are the event yields of $B^0$ and $B_s^0$ mesons extracted from the fit of the two peaks. Possible impact from the peaking backgrounds was estimated with the help of Monte-Carlo simulation and subtracted from the obtained yield values. The branching fractions of the vector mesons decays $\mathcal{B}(K^{*0}\to K^{\pm}\pi^{\pm})$ and $\mathcal{B}(\phi\to K^{+}K^{-})$ were taken as the PDG average \cite{bib:pdg}. The ${f_s}$ and ${f_d}$ stand for the probabilities for a $b$-quark to hadronize into either a $B_s^0$ or $B^0$ meson, the value of their ratio $\frac{f_s}{f_d} = 0.267^{+0.021}_{-0.02}$ measured by LHCb \cite{bib:fsfd} was used here. The last term in the expression $\frac{\varepsilon_{B_s^0\to \phi\gamma}}{\varepsilon_{B^0\to K^{*0}\gamma}}$ is the ratio of the efficiencies, which is actually represented by three terms:
$$\frac{\varepsilon_{B_s^0\to \phi\gamma}}{\varepsilon_{B^0\to K^{*0}\gamma}} =
\frac{\varepsilon^{geo}_{B_s^0\to \phi\gamma}}{\varepsilon^{geo}_{B^0\to K^{*0}\gamma}}
\times\frac{\varepsilon^{reco\&sel}_{B_s^0\to \phi\gamma}}{\varepsilon^{reco\&sel}_{B^0\to K^{*0}\gamma}}
\times\frac{\varepsilon^{trig}_{B_s^0\to \phi\gamma}}{\varepsilon^{trig}_{B^0\to K^{*0}\gamma}},$$
corresponding to the geometrical acceptance of the detector, reconstruction and selection efficiencies and trigger efficiencies respectively. The geometrical acceptance and reconstruction efficiencies are estimated with the help of Monte-Carlo simulation. The selection efficiency is also evaluated from the Monte-Carlo, besides the particle identification efficiencies, which can be estimated directly from the data with the help of calibration samples. The trigger efficiencies are obtained from the data.

\section{Results and conclusions}
In $340~pb^{-1}$ of $pp$ collisions at a center of mass energy of $\sqrt{s} = 7~TeV$ the ratio of branching fractions 
between the $B^0\to K^{*0}\gamma$ and $B_s^0\to\phi\gamma$ channels has been measured to be:
$$\frac{\mathcal{B}(B^0\to K^{*0}\gamma)}{\mathcal{B}(B_s^0\to \phi\gamma)} = 1.52\pm0.14(stat.)\pm0.10(syst.)\pm0.12(f_s/f_d).$$

Combining this ratio of branching fractions with the World Average measurement for the $B^0\to K^{*0}\gamma$, a value for $\mathcal{B}(B_s^0\to \phi\gamma)$ was obtained with an uncertainty much smaller than the previous Belle measurement:
$$\mathcal{B}(B_s^0\to \phi\gamma) = (2.8\pm0.5)\times10^{-5}.$$

The $\mathcal{B}(B_s^0\to \phi\gamma)$ and $\frac{\mathcal{B}(B^0\to K^{*0}\gamma)}{\mathcal{B}(B_s^0\to \phi\gamma)}$ measurements are currently the world's most precise measurement of these quantities and both are compatible with the Standard Model predictions within 1.5 $\sigma$.

\end{document}